\documentclass[11pt,twoside]{article}

\usepackage{asp2014}
\usepackage{scrextend}
\usepackage{xcolor}                  % Good fonts
\usepackage{graphicx}

\aspSuppressVolSlug
\resetcounters

\begin{document}

\title{The CASA software for radio astronomy:\\
status update from ADASS 2019}

\author{B.~Emonts$^{1*}$, R. Raba$^{1*}$, G. Moellenbrock$^{2*}$, S. Castro$^{3*}$, C.E. Garc\'{i}a-Dab\'{o}$^{3*}$, J. Donovan Meyer$^{1}$, P. Ford$^{2}$, R. Garwood$^{1}$, K. Golap$^{2}$, J. Gonzalez$^{3}$, W. Kawasaki$^{4}$, A. McNichols$^{1}$, D. Mehringer$^{1}$, R. Miel$^{4}$, F. Montesino Pouzols$^{3}$, T. Nakazato$^{4}$, S. Nishie$^{4}$, J. Ott$^{2}$, D. Petry$^{3}$, U. Rau$^{2}$, C. Reynolds$^{1}$, D. Schiebel$^{1}$, N. Schweighart$^{1}$, J.-W. Steeb$^{1}$, V. Suoranta$^{1}$, T. Tsutsumi$^{2}$, A. Wells$^{1}$, S. Bhatnagar$^{2}$, $\&$ P. Jagannathan$^{2}$ (o/b/o ARDG), Joe Masters$^{1*}$ (o/b/o pipeline team), K.-S. Wang$^{5*}$ (o/b/o CARTA).
} 
  \affil{$^{1}$NRAO, 520 Edgemont Rd, Charlottesville, VA 22903 \email{casa-feedback@nrao.edu}}
%  \affil{$^{2}$CASA User Liaison \email{casa-feedback@nrao.edu}}  
  \affil{$^{2}$NRAO, 1003 Lopezville Rd, Socorro, NM 87801, USA}
  \affil{$^{3}$ESO, Karl Schwarzschild Strasse 2, D-85748 Garching, Germany}
  \affil{$^{4}$NAOJ, 2-21-1 Osawa, Mitaka, Tokyo 181-8588, Japan}
  \affil{$^{5}$ASIAA, Academia Sinica, PO Box 23-141, Taipei 10617, Taiwan}
%  \affil{$^{7}$IDIA, University of Cape Town, Rondebosch, 7701, South Africa}
%  }
 
\newcommand\blfootnote[1]{%
  \begingroup
  \renewcommand\thefootnote{}\footnote{#1}%
  \addtocounter{footnote}{-1}%
  \endgroup
}

\blfootnote{* Participants ADASS 2019}

\begin{abstract}
CASA, the Common Astronomy Software Applications package, is the primary data processing software for the Atacama Large Millimeter/submillimeter Array (ALMA) and NSF's Karl G. Jansky Very Large Array (VLA), and is frequently used also for other radio telescopes. The CASA software can process data from both single-dish and aperture-synthesis telescopes, and one of its core functionalities is to support the data reduction and imaging pipelines for ALMA, VLA and the VLA Sky Survey (VLASS). CASA has recently undergone several exciting new developments, including an increased flexibility in Python (CASA 6), support of Very Long Baseline Interferometry (VLBI), performance gains through parallel imaging, data visualization with the new Cube Analysis Rendering Tool for Astronomy (CARTA), enhanced reliability and testing, and modernized documentation. These proceedings of the 2019 Astronomical Data Analysis Software $\&$ Systems (ADASS) conference give an update of the CASA project, and detail how these new developments will enhance user experience of CASA.
\end{abstract}

% These lines show examples of subject index entries. At this stage these have to commented
% out, and need to be on separate lines. Eventually, they will be automatically uncommented
% and used to generate entries in the Subject Index at the end of the Proceedings volume.
% Don't leave these in! - replace them with ones relevant to your paper.
%\ssindex{FOOBAR!conference!ADASS 2019}
%\ssindex{FOOBAR!organisations!ASP}

% These lines show examples of ASCL index entries. At this stage these have to commented
% out, and need to be on separate lines. Eventually, they will be automatically uncommented
% and used to generate entries in the ASCL Index at the end of the Proceedings volume.
% The ascl.py command will scan your paper on possible code names.
% Don't leave these in! - replace them with ones relevant to your paper.
%\ooindex{FOOBAR, ascl:1101.010}

\section{Introduction}

The Common Astronomy Software Applications, or CASA \citep[][CASA team et al. in prep.]{mcm07}, is being developed with the primary goal of supporting the data reduction and analysis needs of ALMA and VLA, with a versatility that also benefits the processing of data from other radio telescopes. The CASA infrastructure consists of a set of C++ tools bundled together under an iPython interface as data reduction tasks. This provides flexibility to process the data via task interface or as a Python script, with many post-processing applications available for even more flexibility.

A core aspect of CASA development is support of the ALMA, VLA and VLASS pipelines. The implementation of CASA algorithms and processing techniques that optimize pipeline processing also adds to the continuous improvements made for manual calibration and imaging in CASA. The general user community recommended in the 2018 CASA User Survey to make CASA more reliable \citep{emo18a}.\footnote{CASA Memo $\#$6: https://casa.nrao.edu/casadocs/casa-5.6.0/memo-series/casa-memos/} To honor this request, the CASA team is adopting a new approach to testing, validating and documenting CASA development, which will take shape over the coming CASA releases.

These proceedings provide a status update of the CASA software and highlight a few recent developments that were presented at the 29th Astronomical Data Analysis Software $\&$ Systems (ADASS) conference in Groningen NL, from 6$-$10 Oct. 2019.

\section{Imaging in {\sc tclean}: functionality, speed and reliability}

{\sc tclean} is the CASA task that is used for imaging and deconvolution \citep{rau18}. {\sc tclean} is the successor of {\sc clean}, which is no longer being maintained.

{\sc tclean} has seen recent improvements in mosaicing, widefield and wideband imaging, weighting, deconvolution techniques, and automated masking. In our continuing push to make CASA more reliable for users,$^{1}$ a suite of more than 30 functional verification tests have been written for tclean to evaluate various imaging modes related to joint mosaicing and wideband imaging. The CASA team aims to make certain verification tests for both imaging and calibration available to users in future CASA versions.

CASA now also supports parallel imaging using multiple cores as a standard mode of operation \citep{cas17,cas19}. Speed-up factors of 4-5 can typically be reached for {\sc tclean} on machines with 8-10 cores and sufficient random-access memory \citep{bha15,emo18b}. CASA also improved its memory\,handling for better performance. Parallel imaging can be invoked in the mpicasa environment on normal MeasurementSets. ALMA adopted this mode as default in the Cycle 6 imaging pipeline.

\articlefigure[width=0.93\textwidth]{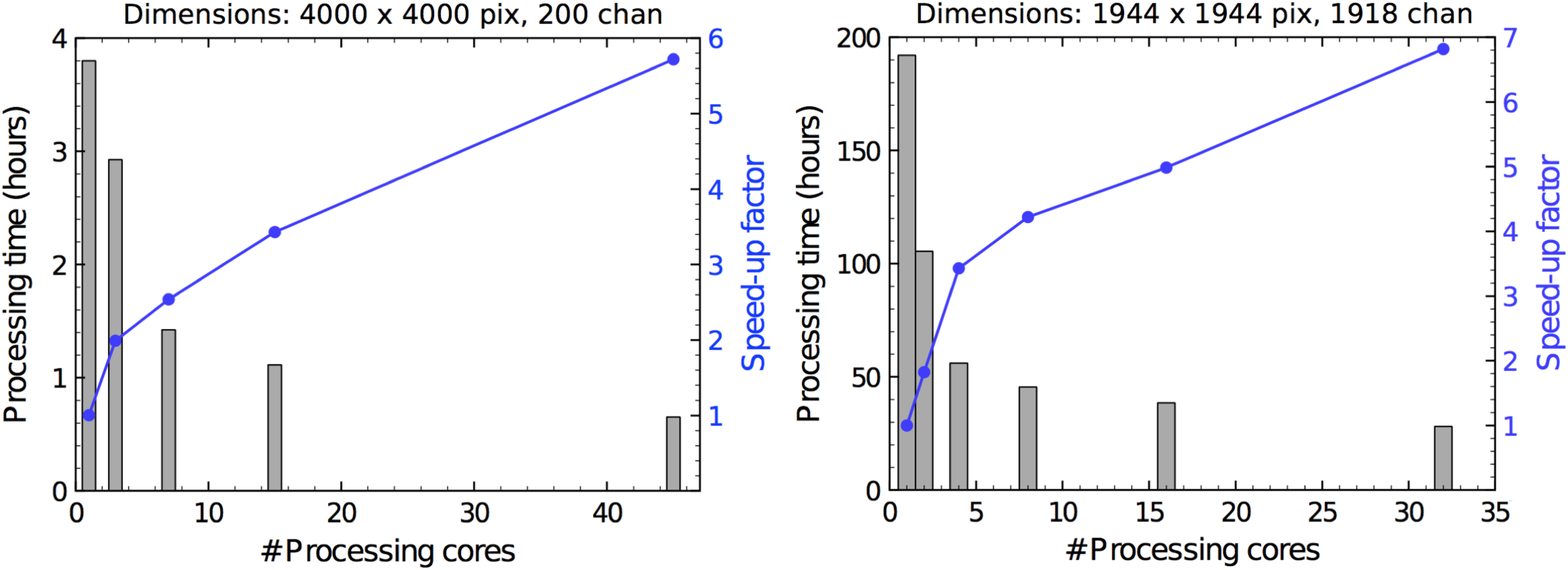}{fig1}{Processing time (left axis) and speed-up factor (right axis) for parallel imaging in {\sc tclean} of two mosaic data sets using multiple cores. The data were processed using the high-performance computing cluster of the North American ALMA Science Center (NAASC). Figure reproduced from \citet{emo19}.}

\section{Very Long Baseline Interferometry}
\label{fig:VLBI}

CASA is also offering increased VLBI support through a collaboration with the Joint Institute for VLBI ERIC (JIVE). JIVE has developed the new CASA tasks {\sc fringefit} for fringe fitting and {\sc accor} for re-normalizing visibilities by their auto-correlation amplitudes \citep[][see also ADASS contribution by Small et al.]{bem19}. Other tasks have also been upgraded to work properly with VLBI observations. An example of how these new VLBI capabilities are powerful tools for the general community is the external VLBI Radboud Pipeline for the Calibration of high Angular Resolution Data (rPICARD), which was developed by \citet{jan19}. This CASA-based VLBI pipeline played a critical role in verifying the calibration of data from the Event Horizon Telescope \citep{eht19,god19}. The CASA team acknowledges JIVE's continuing investment in CASA, and we expect to continue and broaden support for general VLBI processing in the coming years.

\section{CASA 6: Modular Integration in Python}
\label{sec:casa6}

CASA has always been distributed as a single, integrated application, including a Python interpreter and all the libraries, packages and modules. CASA 6 is reorganized to offer a modular approach, where users have the flexibility to build CASA tools and tasks in their Python environment (see ADASS contribution by Raba et al.).

From CASA 6 onward, the tools and tasks are standard Python 3 modules that can be installed via pip wheels. Graphical functionality, such as PlotMS, CASA Viewer, and CARTA will run as separate executable processes. CASA 6 is compatible with Google Colab and Jupyter Notebooks. For details, see \textcolor{blue}{https://go.nrao.edu/casa6}.

Each new CASA release will also contain a monolithic package that is built from the same modular pip wheels, but with a CASA-shell component included. This all-inclusive package replicates the appearance and usage of historic CASA versions.

\section{Data Visualization: CARTA}
\label{sec:carta}

For visualizing image products, most CASA users rely on the CASA Viewer. However, the CASA Viewer is ageing, and the increasing sizes of image cubes become ever more challenging for current visualization tools. The Cube Analysis and Rendering Tool for Astronomy (CARTA; \citealt{com19})\footnote{https://cartavis.github.io} is a new visualization tool designed for ALMA, VLA, MeerKAT, and future radio telescopes, such as the Square Kilometre Array (SKA) and Next-Generation VLA (ngVLA). The CARTA architecture is suitable for visualizing large image cubes through tile rendering and optimized memory use. CARTA is progressing steadily and version 1.2 has been released. New features include region support, tile rendering, efficient spectral display, customizable layouts and HDF5 support. CARTA is expected to replace the CASA Viewer in the foreseeable future.

CARTA is being developed by a team consisting of members from the Academia Sinica Institute of Astronomy and Astrophysics (ASIAA) in Taiwan, the Inter-University Institute for Data Intensive Astronomy (IDIA) in South Africa, the National Radio Astronomy Observatory (NRAO) in the US, and the University of Alberta in Canada.

\section{CASA Documentation and other resources}

CASA Docs is the official online CASA documentation, which has been improved with a more modern look and functionality. CASA Docs offers general information on data processing, as well as detailed task descriptions. The task information also contains parameter descriptions, which can optionally be called inside CASA with the command {\sc doc(`taskname')}. We hope that the improved CASA Docs will enhance user experience.

In addition to the official CASA Docs documentation, there are CASA Guides that provided step-by-step instructions on how to process data from various telescopes. We welcome everyone to use the available resources below to get the most out of CASA.\\
\vspace{-2mm}\\
{\bf CASA Docs} $-$ the official CASA documentation: \textcolor{blue}{https://casa.nrao.edu/casadocs} \\
\vspace{-2.3mm}\\
{\bf CASA Guides} $-$ step-by-step tutorials for data processing: \textcolor{blue}{https://casaguides.nrao.edu}\\
\vspace{-2.3mm}\\
{\bf CASA Website} $-$ official CASA website: \textcolor{blue}{https://casa.nrao.edu} \\
\vspace{-2.3mm}\\
{\bf CASA Newsletter} $-$ twice a year: \textcolor{blue}{https://science.nrao.edu/enews/casa$\_$009}\\
\vspace{-2.3mm}\\
{\bf CASA email lists} $-$ please register: \textcolor{blue}{https://casa.nrao.edu/mail$\_$list.shtml}\\
\vspace{-2.3mm}\\
{\bf Data Processing Questions} $-$ Helpdesks: \textcolor{blue}{https://casa.nrao.edu/help$\_$desk$\_$all.shtml} \\
\vspace{-2.3mm}\\
{\bf CASA feedback} $-$ we welcome general feedback from users! \textcolor{blue}{casa-feedback@nrao.edu}\\

\acknowledgements The CASA team thanks the ADASS 2019 organizers, and the ARDG, pipeline, helpdesk, and CARTA teams for their dedication. CASA is developed by an international consortium of scientists based at the National Radio Astronomical Observatory (NRAO), the European Southern Observatory (ESO), the National Astronomical Observatory of Japan (NAOJ), the Academia Sinica Institute of Astronomy and Astrophysics (ASIAA), the CSIRO division for Astronomy and Space Science (CASS), and the Netherlands Institute for Radio Astronomy (ASTRON), under the guidance of NRAO. The National Radio Astronomy Observatory is a facility of the National Science Foundation operated under cooperative agreement by Associated Universities, Inc. 
%ALMA is a partnership of ESO (representing its member states), NSF (USA) and NINS (Japan), together with NRC (Canada), NSC and ASIAA (Taiwan), and KASI (Republic of Korea), in cooperation with the Republic of Chile. The Joint ALMA Observatory is operated by ESO, AUI/NRAO and NAOJ.  

\bibliography{P5-5_astroph.bib}  % For BibTex

% if we have space left, we might add a conference photograph here. Leave commented for now.
% \bookpartphoto[width=1.0\textwidth]{foobar.eps}{FooBar Photo (Photo: Any Photographer)}

\end{document}